\documentclass[aps, prl, reprint, superscriptaddress]{revtex4-2}

\usepackage[utf8]{inputenc}
\usepackage[T1]{fontenc}
\usepackage{newtxtext}
\usepackage{newtxmath}
\usepackage[scaled=0.92]{helvet}
\usepackage{babel}
\usepackage{microtype}
\usepackage{xspace}
\usepackage{amsmath, amsfonts}
\usepackage{bm}
\renewcommand{\epsilon}{\varepsilon}
\renewcommand{\theta}{\vartheta}
\renewcommand{\rho}{\varrho}
\renewcommand{\phi}{\varphi}

\newcommand*\dif{\mathop{}\!\mathrm{d}}
\renewcommand{\vec}[1]{\bm{#1}}
\newcommand*{\hamil}{\mathcal{H}}
\usepackage[range-units=single, list-units=single, detect-all]{siunitx}
\DeclareSIUnit\rydberg{Ry}
\usepackage{graphicx}
\usepackage[autostyle]{csquotes}
\usepackage[breaklinks, hidelinks]{hyperref}

\newcommand*{\crystaldir}[1]{$[#1]$}

\newcommand*{\ppo}{p_{\parallel\perp}}
\newcommand*{\poo}{p_{\perp\perp}}

\providecommand*{\Fig}{Fig.\@\xspace}
\providecommand*{\FFig}{Figure\@\xspace}

\providecommand*{\eq}{Eq.\@\xspace}

\begin{document} 

\title{Ultrafast coherent all-optical switching of an antiferromagnet with the inverse 
Faraday effect} 

\author{Tobias Dannegger} 
\affiliation{Fachbereich Physik, Universität Konstanz, D-78457 Konstanz, 
Germany} 
\author{Marco Berritta} 
\affiliation{Department of Physics and Astronomy, Uppsala University, Box 516, 
SE-75120 Uppsala, Sweden} 
\author{Karel Carva} 
\affiliation{Charles University, Faculty of Mathematics and Physics, Department 
of Condensed Matter Physics, Ke Karlovu 5, CZ-121 16 Prague, Czech Republic} 
\author{Severin Selzer} 
\affiliation{Fachbereich Physik, Universität Konstanz, D-78457 Konstanz, 
Germany} 
\author{Ulrike Ritzmann} 
\affiliation{Department of Physics and Astronomy, Uppsala University, Box 516, 
SE-75120 Uppsala, Sweden} 
\affiliation{Dahlem Center of Complex Quantum Systems and Department of 
Physics, Freie Universität Berlin, Arnimallee 14, D-14195 Berlin, Germany} 
\author{Peter M. Oppeneer} 
\affiliation{Department of Physics and Astronomy, Uppsala University, Box 516, 
SE-75120 Uppsala, Sweden} 
\author{Ulrich Nowak} 
\affiliation{Fachbereich Physik, Universität Konstanz, D-78457 Konstanz, 
Germany} 

\date{\today} 

\begin{abstract} 
We explore the possibility of ultrafast, coherent all-optical magnetization switching in antiferromagnets by studying the action of the inverse Faraday effect in CrPt, an easy-plane antiferromagnet. Using a combination of density functional theory and atomistic spin dynamics simulations, we show how a circularly polarized laser pulse can switch the order parameter of the antiferromagnet within a few hundred femtoseconds. This nonthermal switching takes place on an elliptical path, driven by the staggered magnetic moments induced by the inverse Faraday effect and leading to reliable switching between two perpendicular magnetic states. 
\end{abstract} 

\maketitle

Since the first observations of laser-induced subpicosecond demagnetization 
phenomena~\cite{Beaurepaire1996}, ultrafast magnetization dynamics have sparked great 
interest, motivated both by the promise of fundamental insights into the physics of ultrafast phenomena as well as potential for data storage applications. 
Purely optically induced magnetization reversal was first observed in the ferrimagnet GdFeCo~\cite{Stanciu2007}. 
The optomagnetic inverse Faraday effect (IFE), in which a magnetic moment is induced by circularly polarized radiation, has been suggested as a possible cause of this effect~\cite{Stanciu2007, Vahaplar2009, Vahaplar2012}. 
However, as was later discovered, \emph{thermally induced} demagnetization is sufficient to achieve switching~\cite{Radu2011, Mentink2012, Ostler2012, Wienholdt2013}. 
In such a helicity-independent process, the magnetization reversal is facilitated by the vastly different speeds with which the sublattices demagnetize~\cite{Radu2011}. 
In ferromagnets, conversely, helicity-dependent switching with the IFE requires a cumulative multishot procedure, like in FePt~\cite{Lambert2014, Hadri2017, John2017}, in which the magnetization is thermally quenched and the IFE sets the direction of remagnetization. These observations have, to varying degree, been ascribed as well to an effect of magnetic circular dichroism on the thermal demagnetization~\cite{Khorsand2012, Ellis2016}. 

While the focus has long been on ferri- and ferromagnets~\cite{Eimuller2001, Lambert2014, Mangin2014, Ellis2015, Nieves2016, Gerlach2017, Hadri2017, John2017, Khorsand2012, Ellis2016}, more recent research has singled out antiferromagnets as interesting alternatives for spintronics applications~\cite{Jungwirth2016, Jungwirth2018}. 
Despite its zero net magnetization, the magnetic state of an 
antiferromagnet can be detected by utilizing, e.g., the anisotropic 
magnetoresistance effect~\cite{Fina2014, Jungwirth2016}. Meanwhile, 
antiferromagnets come with advantages for ultrafast spintronics applications, 
like faster spin dynamics and insensitivity to external magnetic 
fields~\cite{Jungwirth2016, Wunderlich2016}. 

In antiferromagnets, two of the switching mechanisms described above can be ruled out as the sublattices demagnetize with the same speed due to their perfect symmetry, and the magnetic circular dichroism is zero. 
At the same time, the exceptional speed of antiferromagnetic spin dynamics make them more susceptible to ultrashort stimuli like the IFE-induced magnetic moments, which opens up the possibility of a \emph{coherent} switching process triggered by the IFE, which does not involve the temporary demagnetization of the material. 

Here, we assess the plausibility of such a process by calculating the induced magnetic moments \textit{ab initio} using density functional theory (DFT) and then using these results in atomistic spin dynamics simulations to determine the switching probability. 
Previous experiments have already shown that the ultrafast IFE can induce coherent magnon excitations in antiferromagnets such as DyFeO$_3$~\cite{KimelNature2005, Perroni2006}, FeBO$_3$~\cite{Kalashnikova2007}, and NiO~\cite{Satoh2010}. 
In order to achieve an actual switching process, apart from a sufficiently high intensity and short duration of the laser pulse, it is needed that the employed material exhibits a strong IFE and is not very susceptible to laser-induced heating, as that interferes with the coherent switching process as we will show in the following. 
It is also advantageous when the material's anisotropy energy is not so high as to entirely prevent the rotation of the magnetization. 
For detection through anisotropic magnetoresistance, furthermore, one would optimally choose a material with two perpendicular stable magnetization states. 
\begin{figure} 
\includegraphics{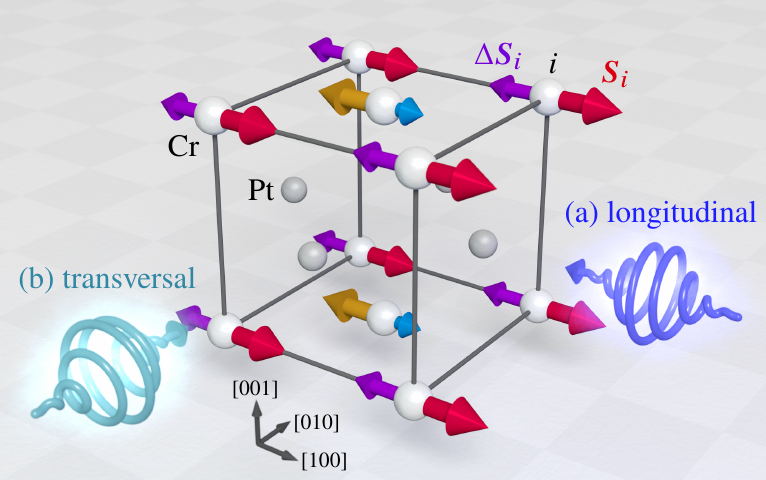} 
\caption{Structure of the CrPt lattice. 
For a longitudinal laser pulse~(a), a magnetic moment $\vec{\mu}_{\text{ind}, 
i}$, corresponding to a spin $\Delta\vec{S}_i$, is induced at each lattice site 
by the IFE (not shown to scale here). 
In the transversal geometry~(b), the induced moments are orders of magnitude 
smaller and are not shown in this figure.} 
\label{fig:Geometry} 
\end{figure} 
Here, we use CrPt as an example system, an easy-plane antiferromagnet in L1$_0$ phase 
with lattice parameters $a = \SI{3.822}{\angstrom}$ and $c = 
\SI{3.811}{\angstrom}$~\cite{Zhang2012} (cf.\ \Fig~\ref{fig:Geometry}). Only 
the Cr atoms possess a significant magnetic moment. 

\begin{figure} 
\includegraphics{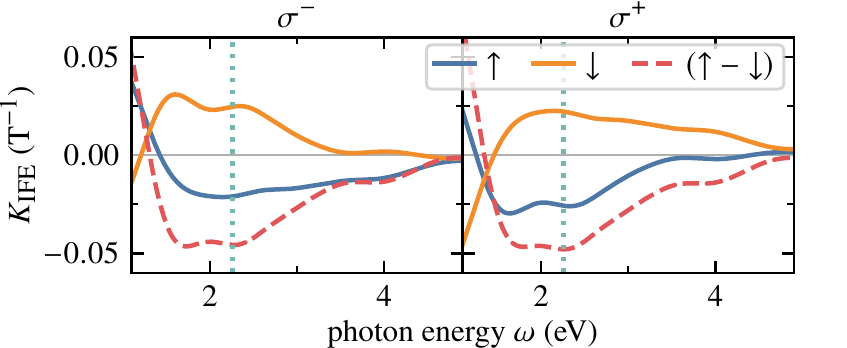} 
\caption{\textit{Ab initio} calculated IFE constants of CrPt, broken down by helicity and 
Cr moment direction (parallel $\uparrow$ and antiparallel $\downarrow$ to the 
$\vec{k}$-vector of the incident light). The dashed red line indicates the 
difference between both directions. For the simulations, a photon energy of 
\SI{2.26}{\electronvolt} was used (marked by the dotted turquoise line).} 
\label{fig:ifecrpt} 
\end{figure} 
We calculate the magnetic moments induced by the IFE \textit{ab initio} using the DFT 
framework developed by Berritta \textit{et al.}~\cite{Marco2016}. 
The induced magnetic moment per spin is given by 
$\mu_\text{ind}^{\uparrow\downarrow} = K_\text{IFE}^{\sigma\uparrow\downarrow}(\omega) V I(t) / c $ 
where $V$ is the volume of each spin moment, $c$ is the speed of light, $I(t)$ is the absorbed intensity of the laser pulse, and $K_\text{IFE}^{\sigma\uparrow\downarrow}(\omega)$ is a constant depending on the photon energy $\omega$, helicity $\sigma$ and the local orientation of the Cr moments, denoted by $\uparrow$ and $\downarrow$. 
The results of our IFE calculations for CrPt in longitudinal geometry are shown in \Fig~\ref{fig:ifecrpt}. 
Interestingly, we find that the induced moments on the Cr sites have a \emph{staggered} symmetry in the sense that they have opposite sign on the two magnetic sublattices. 
In analogy to the staggered magnetization of an antiferromagnet, one can define a staggered induced moment as the difference between the moments induced on the different sublattices. 
This is shown by the dashed line in \Fig~\ref{fig:ifecrpt}. 
The negative sign implies that the induced moments are antiparallel to the local moment on each Cr site. 
The overall effect of the IFE should be maximal at the point where the magnitude of this staggered induced moment is maximal. 
This is the case at a photon energy of \SI{2.26}{\electronvolt}, where the IFE parameters for the two sublattices are 
$K_\text{IFE}^\uparrow = \SI{-0.026}{\per\tesla}$ 
and $K_\text{IFE}^\downarrow = \SI{0.022}{\per\tesla}$, 
for the right-handed ($\sigma^+$) polarization. Due to symmetry breaking, the induced moment is slightly larger on the Cr sublattice that has magnetic moments parallel to the $\vec{k}$-vector than on the other sublattice with antiparallel moments. 
With left-handed polarization it would be the other way around. 
Note that the IFE also induces small moments on the nonmagnetic Pt atoms, but these are not staggered and therefore their effect cancels out~\cite{SupplementalMat}.

We describe the spin system in the classical Heisenberg 
model, where each Cr atom is represented by a normalized spin $\vec{S}_i$. 
The Hamiltonian of the system is 
\begin{equation} 
\hamil = -\sum_{i \neq j} J_{ij} \vec{S}_i \cdot \vec{S}_j 
- \sum_i [ d_z S_{i,z}^2  + 4 d_\text{ip} S_{i,x}^2 S_{i,y}^2 ] 
. 
\end{equation} 
The exchange interaction between the atomic spins is described by the coupling 
energies $J_{ij}$, calculated \textit{ab initio} employing DFT~\cite{Liecht1984, Turek97, Turek2006}. 
The anisotropy term contains a negative uniaxial anisotropy 
$d_z = \SI{-0.61}{\milli\electronvolt}$~\cite{Zhang2012} 
along the \crystaldir{001}-axis, as well as a weaker, fourth-order anisotropy $d_\text{ip} = d_z / 4$ 
that leads to four stable orientations of the Néel vector within the easy plane. 
The coordinate system is chosen such that its axes coincide with the crystallographic axes shown in \Fig~\ref{fig:Geometry}. 

The dynamics of the spin system are computed using the stochastic 
Landau-Lifshitz-Gilbert (LLG) equation~\cite{LandauLifshitz1935, Gilbert2004, 
Brown1963, Nowak2007}: 
\begin{equation} 
\frac{\dif}{\dif t} \vec{S}_i = - \frac{\gamma}{(1+\alpha^2) \mu_s} \vec{S}_i 
\times \bigl( \vec{H}_i + \alpha \vec{S}_i \times \vec{H}_i \bigr) 
  . \label{eq:llg} 
\end{equation} 
Here, $\gamma = \SI{1.76e11}{\per\tesla\per\second}$ is the gyromagnetic ratio, 
$\mu_s = \num{2.24} \mu_B$~\cite{Zhang2012} the saturation moment, 
and $\alpha = 0.05$ the dimensionless Gilbert damping parameter (cf.~\cite{SupplementalMat} for details about the choice of $\alpha$ and the effect of varying it). 
$\vec{H}_i$ is the effective field, given by 
\begin{equation} 
  \vec{H}_i(t) = \vec{\zeta}_i(t) - \frac{\partial\hamil}{\partial\vec{S}_i} 
\end{equation} 
with a random thermal noise $\vec{\zeta}_i(t)$. The first term in 
\eq~\eqref{eq:llg} describes a precession around the effective field and the 
second term leads to a relaxation of the spins towards the effective field. 

We describe the thermal effect of the laser pulse with a simple two-temperature 
model~\cite{KaganovTTM, AnisimovTTM, SupplementalMat}. 
To describe the IFE, a spin $\Delta\vec{S}_i = {\vec{\mu}_{\text{ind}, 
i}}/{\mu_s}$ is added to the spin $\vec{S}_i$ in all the terms of the Hamiltonian. The exchange term can then be rewritten as 
\begin{align} 
\hamil_\text{exc} &= -\sum_{i \neq j} J_{i j} (\vec{S}_i + \Delta\vec{S}_i) 
                     \cdot (\vec{S}_j + \Delta\vec{S}_j) \nonumber\\ 
  &= -\sum_{i \neq j} J_{i j} \vec{S}_i \cdot \vec{S}_j  
     - 2 \sum_{i \neq j} J_{i j} \vec{S}_i \cdot \Delta\vec{S}_j + \mathrm{const.} , 
\end{align} 
where the first term describes the regular exchange interaction between 
neighboring spins and the second term describes the interaction of a spin with 
the induced moments on neighboring lattice sites. 
The change of the saturation moment $\mu_s$ in the presence of the induced moments is small enough to be negligible~\cite{SupplementalMat}. 

For the DFT calculations, a stationary state is considered where the light 
intensity remains at a constant value and the induced magnetic 
moments are hence directly proportional to the intensity. The exact time 
dependence of the induced moments in the ultrafast regime, however, is 
unknown but it is likely not directly proportional to the momentary laser 
intensity on subpicosecond timescales~\cite{Popova2011, Vahaplar2012}. We 
therefore assume, as was done before in similar simulations~\cite{John2017}, 
that the induced moments subside exponentially after the laser pulse on a 
certain timescale $\tau$, which was varied between \SIlist{0; 
250}{\femto\second}. The following results, unless specified otherwise, use a 
value of $\tau = \SI{100}{\femto\second}$. 

All simulations were initialized with all spins aligned parallel to the 
crystal's \crystaldir{100}-axis and an initial temperature of \SI{300}{\kelvin}. The $\vec{k}$-vector of the incident laser 
beam is oriented along the \crystaldir{\bar{1}00}-direction in the longitudinal 
geometry or along \crystaldir{010}-direction in transversal geometry  
(cf.\ \Fig~\ref{fig:Geometry}). The duration of the laser pulse is always set 
to \SI{60}{\femto\second} (full width at half maximum). 

\begin{figure} 
\includegraphics{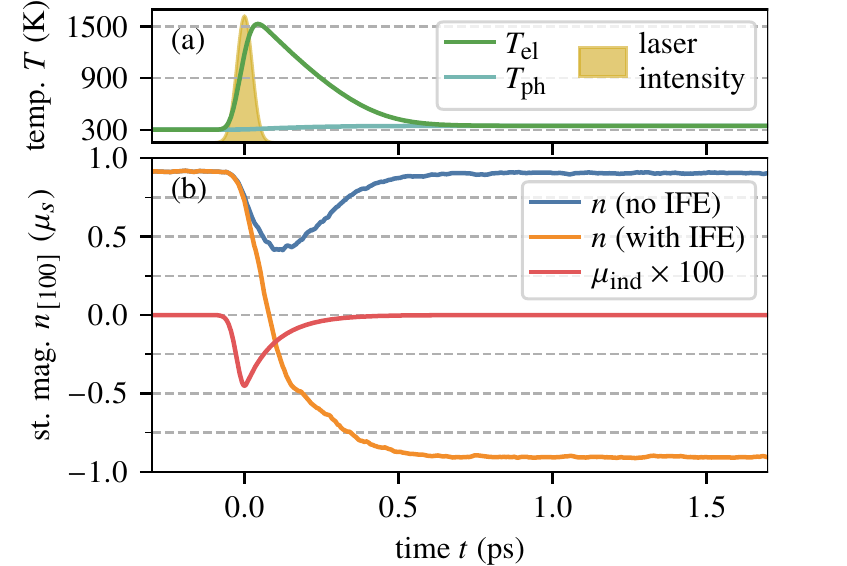} 
\caption{Simulated time evolution of the CrPt system. 
(a) Shape of the laser pulse occurring at time $t = 0$ and resulting 
temperatures of the electronic and phononic heat bath. 
(b) Time evolution of the \crystaldir{100}-component of the staggered magnetization with and without 
the IFE taken into account (blue and orange lines) and staggered induced 
magnetic moment (red line), also normalized to the saturation magnetization but 
magnified by a factor of \num{100}.} 
\label{fig:ExampleSketch} 
\end{figure} 

In the longitudinal configuration, magnetic moments will be induced by the IFE 
that reach between about \SIlist{.1;1}{\percent} of the system's saturation 
magnetization, depending on the laser intensity. \FFig~\ref{fig:ExampleSketch} 
shows the results of an example simulation with an absorbed laser intensity of 
$I = \SI{4.2}{\giga\watt\per\square\centi\metre}$. The electron system heats up 
very quickly as the laser pulse arrives, bringing the spin system far above its 
critical temperature ($T_N \approx \SI{763}{\kelvin}$) before cooling down 
again by dissipating heat to the phonon system. During the high-temperature 
period the system's staggered magnetization diminishes markedly but not below 
approximately half the saturation value. After about half a picosecond, the 
system has reached thermal equilibrium again. 
To show the influence of the IFE, we carried out simulations both with and without the induced magnetic moments. 
Without the IFE, the magnetization of the sublattices is temporarily 
reduced but then remagnetizes without changing its direction. With the IFE, the 
induced magnetic moments exert a torque moving the spins towards the opposite 
direction. This reversal process takes only about \SI{200}{\femto\second}. 
The initial torque is only produced if the induced magnetic moments are not exactly collinear to the magnetization, but laser excitation and thermal fluctuations at room temperature lead to sufficient deviations from the ground state for a significant torque to be exerted. 

This example shows that ultrafast switching via the IFE is possible in CrPt. 
However, the underlying process is not deterministic and the switching 
probability depends on many parameters, like the laser intensity, the 
absorption coefficient, the angle of incidence, the shape of the laser pulse, 
and the way the temporal evolution of the induced magnetic moments is modeled. 
In order to determine the probability of a switching process to occur for a 
certain set of parameters, each simulation was repeated several hundred times 
for different realizations of the thermal noise. 

\begin{figure} 
\includegraphics{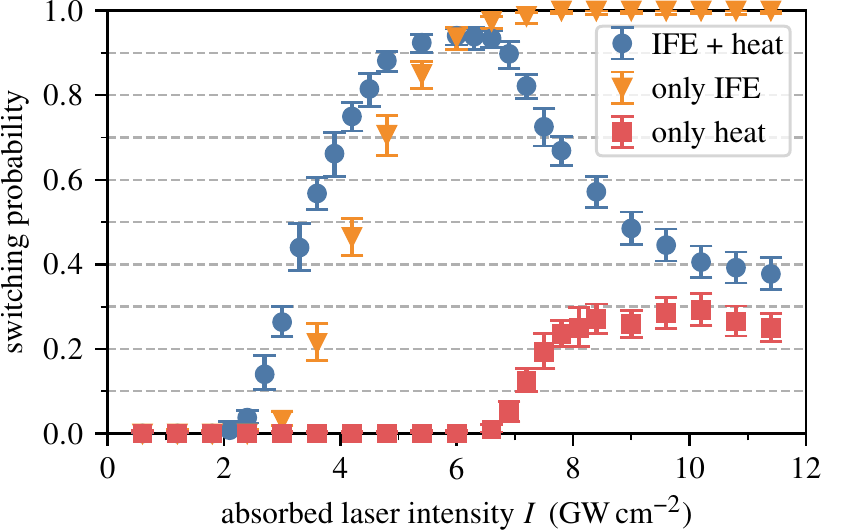} 
\caption{Simulated \SI{180}{\degree} switching probabilities in longitudinal 
geometry. 
The error bars indicate the \SI{95}{\percent} confidence interval.} 
\label{fig:Results} 
\end{figure} 
The statistical results for \SI{180}{\degree} switching in longitudinal 
geometry are presented in \Fig~\ref{fig:Results}. 
To discern the dissipative from the non-dissipative effects, simulations were also performed without the induced magnetic moments (only laser-induced heating) and without the heating (magnetic moments are induced, but the temperature remains at \SI{300}{\kelvin}). 
Without the IFE, the switching probability remains close to zero and only increases for higher laser intensities, where the spin system becomes strongly demagnetized, such that the resulting orientation is randomized. 
Without the heating, the switching probability still goes up to \SI{100}{\percent} and does not decrease for higher intensities. 
This shows that the switching process we observe here is caused by the IFE and cannot be attributed to dissipative effects. 
The switching in CrPt is thus notably distinct from that of ferrimagnetic GdFeCo, which is a thermal process that requires sufficient heat to nearly demagnetize the sublattice moments before the magnetization builds up again~\cite{Radu2011, Ostler2012}. 
Conversely, in CrPt a larger laser heating works against the coherent switching action of the IFE (Fig.~\ref{fig:Results}). 

We have performed similar simulations for an L1$_0$ FePt spin lattice, using the spin 
model developed by Mryasov \textit{et al.}~\cite{Mryasov2005}, so we can compare the 
behavior of antiferromagnetic CrPt to that of ferromagnetic FePt. The 
situation in FePt is different in as much as it is an easy-axis ferromagnet 
with a very strong anisotropy. This allows only for \SI{180}{\degree} switching 
in FePt, along the easy anisotropy axis. Our simulations of the switching 
process in FePt confirm previous findings~\cite{John2017}, according to which 
no deterministic single-pulse switching occurs in FePt. The IFE can only 
slightly increase the switching probability. Hence, in order to change the 
magnetization of an ensemble of FePt nanograins, multiple laser pulses need to 
be applied~\cite{Lambert2014}. 

One factor that makes the switching process easier in CrPt is that the 
anisotropy barrier is much lower than in FePt. In FePt, the laser intensity 
needs to be high enough to completely demagnetize the spin lattice before it 
can then remagnetize in the opposite direction. 
\begin{figure} 
\includegraphics{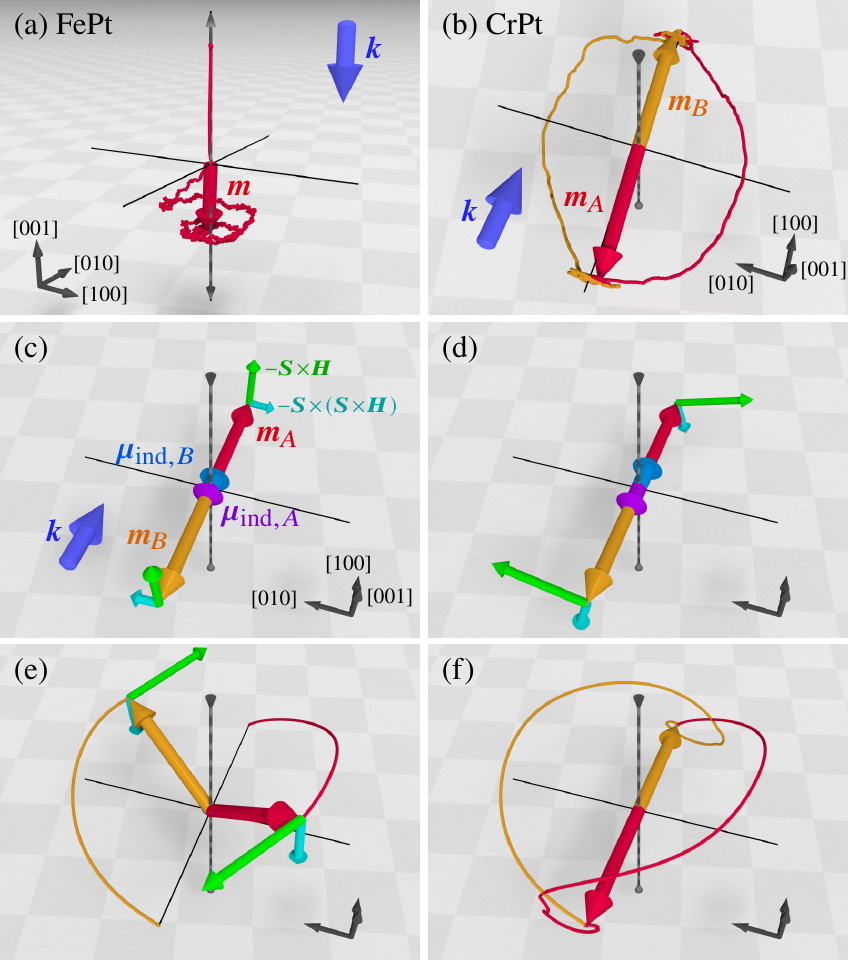} 
\caption{Comparison of the switching paths of (a) the magnetization vector 
$\vec{m}$ of the FePt system and (b) the sublattice magnetization vectors 
$\vec{m}_A$ and $\vec{m}_B$ of the CrPt system during a \SI{180}{\degree} 
switching process. Panels (c) -- (f) show a schematic depiction of the exchange-enhanced 
antiferromagnetic switching process without thermal noise. The precession and 
damping torque are represented by the green and cyan arrows, respectively, 
where the damping torque has been magnified by a factor of 10 relative to the 
precession torque to keep it visible. For clarity, the canting angle between 
the sublattices has also been exaggerated by a factor of 10.} 
\label{fig:Trajectories} 
\end{figure} 
The trajectory of the magnetization vector during this process is illustrated 
in \Fig~\ref{fig:Trajectories}(a). The time it takes the spin system to fully 
demagnetize is close to a picosecond. After that time, the magnetic moments 
induced by the laser pulse are no longer present. 
Then, the magnetization slowly builds up again, leading to a \emph{linear} reversal path (cf.~\cite{Kazantseva2009}), similar to GdFeCo~\cite{Vahaplar2009}. 
Consequently, we find that the IFE can only have a marginal effect on the remagnetization direction, consistent with experiments \cite{Lambert2014, John2017}. 

We further find that when the CrPt model is artificially made ferrimagnetic by varying the magnetic moment of one sublattice with respect to the other, the switching probability diminishes rapidly (data on this can be found in~\cite{SupplementalMat}). 
In antiferromagnetic CrPt, however, the Néel vector can simply rotate within the easy plane without the material losing its magnetic order [cf.~\Fig~\ref{fig:Trajectories}(b)]. 
The whole rotation only takes a few hundred femtoseconds, not much longer than the laser pulse itself. 

The main reason for this difference in speed is the exchange-enhanced dynamics 
of antiferromagnets. The interplay of damping and precession torque that leads 
to this effect is visualized schematically in 
\Fig~\ref{fig:Trajectories}(c--f). The precession term dominates the dynamics 
on short timescales because the damping term is usually much smaller as it 
scales with the damping parameter $\alpha \ll 1$. At the beginning of the 
switching process [panel (c)] the induced magnetic moments cause the precession 
term of the LLG, which is perpendicular to the effective field, to drive the 
sublattice magnetizations out of the easy plane. But because the moments are 
staggered, the direction of precession is opposite on the two sublattices, such 
that the precessional torque pulls the spins on both sublattices in the same 
direction, leading to a small canting between the sublattice magnetizations. 
This canting then leads to an effective field produced by the exchange 
interaction between the sublattices, which is much stronger than that caused by 
the induced moments. This effective field is oriented along the anisotropy 
axis, such that the following precession around it leads to an in-plane 
rotation of the spins [see panel (d), which shows the moment of maximum laser 
intensity]. Panel (e) shows the situation shortly after the laser pulse. By 
that time, the induced moments are no longer present in the system. But because 
the sublattices are still canted, the switching motion continues driven by the 
precessional torque. This effect is known as inertial switching~\cite{Kimel2009, 
Wienholdt2012}. Because of the comparatively low damping torque, the canting 
subsides only slowly. Finally, panel (f) shows the complete switching path. 

If slightly lower laser intensities are used, the induced moments will be 
weaker and the Néel vector more likely to rotate only \SI{90}{\degree} instead 
of \SI{180}{\degree}. 
\begin{figure} 
\includegraphics{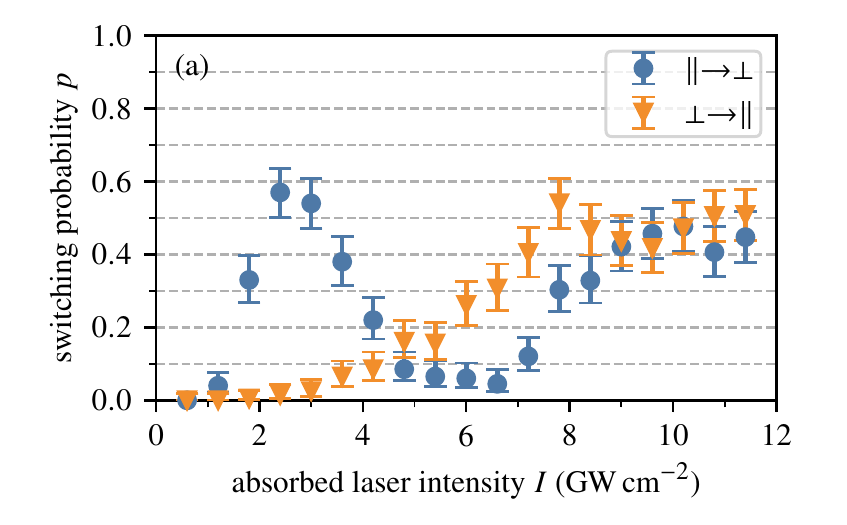}\\ 
\includegraphics{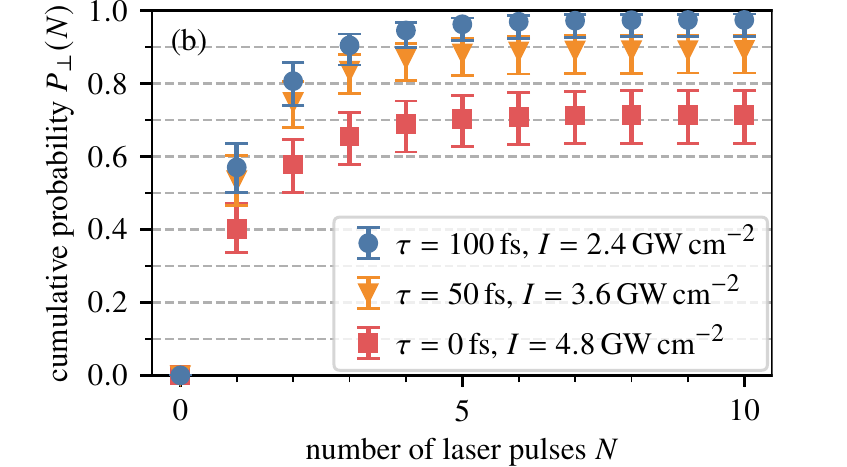} 
\caption{(a) Probability of switching a system from a magnetization parallel to the 
incident laser light to a perpendicular state and vice versa. 
(b) Probability of finding the system in a perpendicular state after $N$ laser 
pulses, when starting with an initial state parallel to the $\vec{k}$-vector. 
The error bars indicate the \SI{95}{\percent} confidence interval.} 
\label{fig:ResultsNinety} 
\end{figure} 
\FFig~\ref{fig:ResultsNinety}(a) shows the computed probability of achieving 
\SI{90}{\degree} switching that way for various laser intensities. This 
probability does not get much higher than \SI{50}{\percent} but, as 
\Fig~\ref{fig:ResultsNinety}(a) also shows, once the system is in a 
perpendicular state, the probability of switching back into a state parallel to 
the $\vec{k}$-vector of the incident light is very low because only negligible 
magnetic moments are induced by the IFE with transversal incidence. 

So if several consecutive laser pulses are used, the amount of switched systems 
in an ensemble will increase with every pulse. If $\ppo$ describes the 
probability of switching from a parallel to a perpendicular state and $1 - 
\poo$ describes the probability of switching back (with $\poo$ being the 
probability of remaining in the perpendicular state after a pulse), then the 
switching probability after infinitely many pulses is given by $P_\perp(\infty) 
= \ppo [1 - (\poo - \ppo)]^{-1}$. The probability of being in a perpendicular 
state after $N$ pulses can be expressed as $P_\perp(N) = P_\perp(\infty) + 
(P_\perp(0) - P_\perp(\infty)) (\poo - \ppo)^N$. 
The results of this calculation are shown in \Fig~\ref{fig:ResultsNinety}(b), for different values of the decay time $\tau$. 
We find that with an experimentally realistic laser intensity it should be 
possible to reach a very high switching probability after only a few pulses. 

In summary, our spin dynamics simulations show how all-optical magnetization 
switching can be induced by the IFE in an easy-plane antiferromagnet. The IFE 
is found to induce staggered moments and to have a much more pronounced effect in the antiferromagnetic spin 
lattice than in ferro- and ferrimagnetic systems. 
It is the particular properties of antiferromagnets that allow for exchange-enhanced, inertial switching on an elliptical path. This process is triggered by staggered magnetic moments induced by the IFE and takes place within a few hundred femtoseconds and, for some parameter values, with close to \SI{100}{\percent} probability. 
Furthermore, our results suggest that 
a sequence of multiple laser pulses can be used for controllable switching 
between two perpendicular magnetization states. The latter result is most 
important for an experimental validation of this switching process since 
\SI{90}{\degree} switching is easily detectable via anisotropic 
magnetoresistance measurements. 

\begin{acknowledgments} 
We thank J.\ Hurst for valuable discussions. 
This work has been supported by the German Research Foundation (DFG) under Grant No.\ 290/5-1 and through CRC/TRR 227, the Swedish Research Council (VR), the K.\ and A.\ Wallenberg Foundation (Grant No.\ 2015.0060), the Horizon2020 Framework Programme of the European Commission under FET-Open Grants No.\ 737093 (FemtoTerabyte) and 863155 (s-Nebula), and the Czech Science Foundation (Grant No.\ 19-13659S). Part of the calculations were enabled by resources provided by the Swedish National Infrastructure for Computing (SNIC) at NSC Linköping partially funded by the Swedish Research Council through grant agreement No.\ 2018-05973, and by the Ministry of Education, Youth and Sports of the Czech Republic through the e-INFRA CZ (ID:90140). 
\end{acknowledgments}

\nocite{John2017, Radu2011, Ostler2012, Papusoi2018, Strungaru2020}

\end{document}